# Effects of numerical implementations of the impenetrability condition on non-linear Stokes flow: applications to ice dynamics

Christian Helanow[1]

[1]Stockholm University, Department of Physical Geography, SE-106 91 Stockholm, Sweden

**Abstract**

The basal sliding of glaciers and ice sheets can constitute a large part of the total observed ice velocity, in particular in dynamically active areas. It is therefore important to accurately represent this process in numerical models. The condition that the sliding velocity should be tangential to the bed is realized by imposing an impenetrability condition at the base. We study the, in glaciological literature used, numerical implementations of the impenetrability condition for non-linear Stokes flow with Navier's slip on the boundary. Using the finite element method, we enforce impenetrability by: a local rotation of the coordinate system (strong method), a Lagrange multiplier method enforcing zero average flow across each facet (weak method) and an approximative method that uses the pressure variable as a Lagrange multiplier for both incompressibility and impenetrability. An analysis of the latter shows that it relaxes the incompressibility constraint, but enforces impenetrability approximately if the pressure is close to the normal component of the stress at the bed. Comparing the methods numerically using a method of manufactured solutions unexpectedly leads to similar convergence results. However, we find that, for more realistic cases, in areas of high sliding or varying topography the velocity field simulated by the approximative method differs from that of the other methods by $\sim 1\%$ (two dimensional flow) and $> 5\%$ when compared to the strong method (three-dimensional flow). In this study the strong method, which is the most commonly used in numerical ice sheet models, emerges as the preferred method due to its stable properties (compared to the weak method in three dimensions) and ability to well enforce the impenetrability condition.

## 1 Introduction

An accurate representation of the dynamics of ice sheets and glaciers is an important component of increasing our understanding about past and future climate. Glaciers and ice sheets currently contribute to sea-level rise (Church *et al.*, 2013) and can influence large scale weather patterns and ocean circulation (Clark *et al.*, 1999), as well as playing vital roles in triggering abrupt climatic events in the past (Heinrich, 1988). The physical domain that glaciers occupy and how much mass they store or release therefore becomes intricately linked to the climate and possible feedback effects (Zhang *et al.*, 2014).

Ice, through a constitutive relation, can be considered to be an incompressible (singular) power-law fluid of very viscous type (Glen, 1955). The dynamics can then, with minimal simplifications, be described as a gravity driven free surface flow governed by the the non-linear Stokes equations. These partial differential equations are considered to be the most accurate representation of the physics of ice deformation, and are often called the *Full Stokes* (FS) equations in the context of the ice modeling community and are solved for the glacier velocity and pressure fields.

The use of numerical models has become an indispensable tool to glaciologists and climate scientists, both to understand paleo-ice sheets and for prognostic simulations of climate. Today, a multitude of models of various complexity exist and multiple of these use the framework of finite element methods (FEM) to solve the FS equations, e.g. Elmer/Ice (Gagliardini *et al.*, 2013), VarGlaS (Brinkerhoff and Johnson, 2013) and ISSM (Larour *et al.*, 2012).

The FS equations describe the flow of ice that is due to internal deformation, however when considering the total velocity distribution of ice, aspects of how the glacier slides over the underlying substrate and how this can deform plays an important role (e.g. Cuffey and Paterson, 2010). It is not uncommon that sliding and substrate deformation dominates the total movement of the ice. For instance, Hooke *et al.* (1997) estimated that the sliding speed at Storglaciären (valley glacier in NW Sweden) accounts for over 85% of the total (surface) velocity, with a similar value given for a land terminating part of the Greenland Ice Sheet (Sole *et al.*, 2013). In general these processes, summed up by the sliding velocity (to be solved for), become a part of the boundary condition *necessary* to close and solve the partial differential equations. In conjunction with the sliding boundary condition, an impenetrability condition is also specified (i.e. the ice cannot penetrate the bed making velocity tangential).

The focus of this study is on different implementations of the impenetrability condition in FEM and if, or how, the velocity and pressure distribution is affected by three different methods; a *strongly imposed*, a *weakly imposed* and an *approximative* method. We use manufactured solutions from Sargent and Fastook (2010) and Leng *et al.* (2013) to investigate the convergence rates and validity of each method, specifically how the solutions generated differ at the bed and surface. We further compare the methods in two dimensions by using a benchmark glacier experiment and contrast the strong and approximative method in a three dimensional simulation of the Greenland Ice Sheet.

The study is structured as follows: Section 2 introduces the equations considered, followed by the implementation of the impenetrability methods in Section 3. We present the results in Section 4 and finally summarize our conclusions in Section 5.

## 2 Governing equations

In this study we will focus on the specific case of a grounded isothermal ice sheet or glacier occupying the domain $\Omega \subset \mathbb{R}^d$ (considered to be polygonal for simplicity), where $d = \{2, 3\}$ is the dimension. The boundary of the domain, $\Gamma$, is divided into that of the free surface (parts of the glacier in contact with the atmosphere, denoted by $\Gamma_s$) and the parts that are in contact with the underlying rigid bed, $\Gamma_b$. As is customary for the free surface of the glacier, wind stresses and atmospheric pressure are neglected resulting in a stress-free surface. At the glacier bed we prescribe two boundary conditions, that of impenetrability and relating the tangential shear stresses to tangential velocities at the bed (basal slip).

The above result in a set of non-linear partial differential equations presented in subsequent sections.

## 2.1 The power-law Stokes equations

In a Cartesian coordinate system $(x, y, z)$ (when $d = 2$, $z$ will be used as the vertical coordinate), the velocity $\mathbf{u} = (u, v, w)$ and pressure $p$ are given by the solution to the power-law Stokes equations

$$-\nabla p + \nabla \cdot \mathbf{S} + \rho \mathbf{g} = \mathbf{0}, \tag{1a}$$
$$\nabla \cdot \mathbf{u} = 0, \tag{1b}$$

where $\rho$ is the ice density, $\rho\mathbf{g}$ the gravitational body force, and $\mathbf{S}$ is the deviatoric stress tensor. Equations (1a) and (1b) describe the balance of momentum and the conservation of mass (incompressibility) respectively. The deviatoric stress tensor is related to the strain-rate tensor $\mathbf{D} = \frac{1}{2}(\nabla\mathbf{u}+\nabla\mathbf{u}^T)$ through a power-law,

$$\mathbf{S}(\mathbf{u}) = 2\eta(\mathbf{II}_{\mathrm{D}})\mathbf{D}, \tag{2}$$

where the ice viscosity $\eta$ is a function of the second invariant of the strain-rate tensor $\mathbf{II}_{\mathrm{D}} = \mathbf{D} : \mathbf{D} = \mathbf{D}_{ij}\mathbf{D}_{ij}$. In glaciology the constitutive relation

$$\eta = \eta_0 \mathbf{II}_{\mathrm{D}}^{(1-n)/2n}, \tag{3}$$

called Glen's flow law, is most commonly used. The power-law parameter $n$ indicates the non-linearity of the material and is taken to equal $n = 3$ (Glen, 1955). Normally $\eta_0$ is a spatially varying and temperature-dependent parameter, however, as mentioned above we consider the isothermal case with $\eta_0$ being constant. Together, (2) and (3) make ice a shear-thinning fluid which, in the limit of $\mathbf{II}_{\mathrm{D}} = 0$, has infinite viscosity.

## 2.2 Boundary conditions

Denoting the boundary of the domain as $\Gamma = \Gamma_s \cup \Gamma_b$, where the indices $s$ and $b$ specify the surface and bed, and the Cauchy (total) stress tensor as $\mathbf{T} = \mathbf{S} - \nabla p$, the stress-free condition at the surface is given by

$$\mathbf{T} \cdot \mathbf{n} = \mathbf{0} \quad \text{on } \Gamma_s, \tag{4}$$

where $\mathbf{n}$ is the outward pointing unit normal. This is a Neumann boundary condition and is naturally included in the weak formulation of (1) (see Section 3.2).

Impenetrability amounts to requiring that the normal component of the velocity at the bed be zero, that is

$$\mathbf{u} \cdot \mathbf{n} = 0 \quad \text{on } \Gamma_b. \tag{5}$$

Equation (5) is a Dirichlet boundary condition in the normal direction of the velocity. Finally, we assume a Navier's slip type boundary condition at the bed

$$\mathbf{t_i} \cdot \mathbf{S} \cdot \mathbf{n} = -\beta^2 \mathbf{u} \cdot \mathbf{t_i} \quad (i = 1, \ldots, d-1), \tag{6}$$

where $\mathbf{t_i}$ is are tangent vectors spanning the plane orthogonal to $\mathbf{n}$ and $\beta^2 \geq 0$ is the basal drag coefficient. For high values of $\beta^2$ little sliding is exhibited while free-slip conditions are present at $\beta^2 = 0$. Note that, since

$$\mathbf{t_i} \cdot \mathbf{T} \cdot \mathbf{n} = \mathbf{t_i} \cdot \mathbf{S} \cdot \mathbf{n} - p\mathbf{t_i} \cdot \mathbf{n} = \mathbf{t_i} \cdot \mathbf{S} \cdot \mathbf{n},$$

we have that (6) is equivalent to relating the tangential components of the Cauchy stress linearly to the basal velocity, $\mathbf{u}_b$. Therefore, the sliding boundary condition is again a Neumann-type boundary condition naturally present in the weak formulation of the Stokes equations. However, it is only to be applied to the tangential components of the velocity at bed.

Herein we only consider the linear relation stated in (6), but it is worth to note that other relations can and have been specified in the field of glaciology. For instance, a Weertman-type "sliding law" relates $\mathbf{u}_b^{\tilde{n}}$ to the basal shear stress, where $\tilde{n}$ is related to $n$ (Weertman, 1957). Even further developments to the sliding law that take into consideration sliding over beds where cavities might form have been made by e.g. Schoof (2005). Ultimately, the sliding is going to be interlinked with thermal and hydrological processes at the bed and how these interact with the underlying substrate.

### 2.3 Weak formulation

Due to the nonlinear nature of (1), the function spaces for the velocity and pressure are related to Glen's parameter $n$. To facilitate notation, we set $\mathfrak{p} = \frac{n+1}{n}$ and $\mathfrak{p}^* = n + 1$ (for the interested reader, the problem is often called a $\mathfrak{p}$-Stokes problem (e.g. Hirn, 2011; Belenki *et al.*, 2012), where $\frac{1}{\mathfrak{p}} + \frac{1}{\mathfrak{p}^*} = 1$). The appropriate spaces are then

$$\begin{aligned} \mathcal{V} &:= \left\{ \mathbf{v} \in [W^{1,\mathfrak{p}}(\Omega)]^d \right\}, \\ \mathcal{Q} &:= L^{\mathfrak{p}^*}(\Omega). \end{aligned}$$

Here, the velocity space is the Sobolev space consisting of functions whose first derivative lies in $L^{\mathfrak{p}}(\Omega)$. The weak formulation is then obtained by multiplying (1a) and (1b) by test functions $\mathbf{v} \in \mathcal{V}$ and $q \in \mathcal{Q}$ respectively and integrating by parts to arrive at

$$\int_\Omega \mathbf{S}(\mathbf{u}) : \nabla \mathbf{v}\, d\Omega - \int_\Omega p \nabla \cdot \mathbf{v}\, d\Omega - \int_{\Gamma_b} \mathbf{n} \cdot (\mathbf{S}(\mathbf{u}) - p\mathbf{I}) \cdot \mathbf{v}\, d\Gamma = \int_\Omega \rho \mathbf{g} \cdot \mathbf{v}\, d\Omega \quad \forall \mathbf{v} \in \mathcal{V}, \tag{7a}$$

$$\int_\Omega \nabla \cdot \mathbf{u} q\, d\Omega = 0 \quad \forall q \in \mathcal{Q}. \tag{7b}$$

Jouvet and Rappaz (2011) shoved that (7) together with the considered boundary conditions (1) and (4) to (6) make the problem well-posed. As can be seen in (7), the pressure variable $p$ is acting as a Lagrange multiplier that enforces the incompressibility condition. We have not yet decided on how to deal with the boundary conditions at the bed, but the stress free condition (4) at the surface has been included above by setting

$$-\int_{\Gamma_s} \mathbf{n} \cdot (\mathbf{S}(\mathbf{u}) - p\mathbf{I}) \cdot \mathbf{v}\, d\Gamma_s = 0.$$

Since the surface boundary condition is present in the weak formulation, the boundary condition is said to be enforced *weakly*. If we would instead restrict the function space used for the test functions $\mathbf{v}$ (or $p$) by specifying a value at the boundary, this condition would be enforced *strongly*. In the upcoming section the focus is on the implementing the impenetrability condition strongly, weakly and in an approximative manner.

## 3 Enforcing impenetrability

For discretization we employ the finite element method. Let $\mathcal{T}_h = \{K\}$ be a partition of the domain $\Omega$ into triangles or tetrahedrons, which are denoted by $K$. The set of facets, $F$, lying on $\Gamma_b$ can then be defined as

$$\mathcal{F}_h = \{F : F \in K \cap \Gamma_b, K \in \mathcal{T}_h\}.$$

As approximations to the function spaces $\mathcal{V}$ and $\mathcal{Q}$, we will use piecewise polynomials defined by

$$\begin{aligned}
\mathcal{V}_h^k &= \{\mathbf{v} \in [C(\overline{\Omega})]^d : \mathbf{v}|_K \in [\mathcal{P}_k]^d, k = 1 \text{ or } 2\}, \\
\mathcal{Q}_h &= \{q \in C(\overline{\Omega}) : q|_K \in \mathcal{P}_1\},
\end{aligned}$$

where $\mathcal{P}_k$ the set of polynomials of degree $k$. For the two-dimensional simulations we have opted to use the Taylor-Hood element ($P2P1$), i.e. $\mathcal{V}_h^2 \times \mathcal{Q}_h$. This since the element pair fulfills the *inf-sup* condition (Babuška, 1973; Brezzi, 1974), avoiding extra stabilization terms. The discrete variational problem then reads: find $(\mathbf{u}_h, p_h) \in \mathcal{V}_h^2 \times \mathcal{Q}_h$ such that

$$\begin{aligned}
\int_\Omega \mathbf{S}(\mathbf{u}_h) : \nabla \mathbf{v}_h \, d\Omega - \int_\Omega p_h \nabla \cdot \mathbf{v}_h \, d\Omega - \int_\Omega \nabla \cdot \mathbf{u}_h q_h \, d\Omega \\
- \int_{\Gamma_b} \mathbf{n}_h \cdot (\mathbf{S}(\mathbf{u_h}) - p_h \mathbf{I}) \cdot \mathbf{v}_h \, d\Gamma \\
= \int_\Omega \rho \mathbf{g} \cdot \mathbf{v}_h \, d\Omega \quad \forall (\mathbf{v}_h, p_h) \in \mathcal{V}_h^k \times \mathcal{Q}_h,
\end{aligned} \tag{8}$$

where $\mathbf{n}_h$ denotes the discrete outward unit normal. In three-dimensions, the above elements were deemed too expensive, and we used the lower order pair $\mathcal{V}_h^1 \times \mathcal{Q}_h$ ($P1P1$). Contrary to $P2P1$, this element needs to be stabilized. This has to do with the fact that (7) is a saddle-point problem resulting from that the incompressibility is enforced by the pressure acting as a Lagrange multiplier. We have approached this problem in a standard, and in glaciology commonly used, way by adding the Galerkin Least-Squares (GLS) terms (e.g. Hughes *et al.*, 1986; Franca and Frey, 1992)

$$\begin{aligned}
&- \sum_{K \in T_h} \tau_{\text{GLS}} \int_K \nabla p_h \cdot \nabla q_h \, dK \qquad \text{(LHS)}, \\
&- \sum_{K \in T_h} \tau_{\text{GLS}} \int_K \rho \mathbf{g} \cdot \nabla q_h \, dK \qquad \text{(RHS)}.
\end{aligned} \tag{9}$$

In this study we have chosen to define the stabilization parameter as

$$\tau_{\text{GLS}} = \tau_0 \frac{m_K h_K^2}{8 \eta_{lin}}, \tag{10}$$

where $m_K = 1/3$ for linear interpolations, $h_K$ is measure of the cell size defined as the *minimum edge length* of the cell, $\eta_{lin} = 10^{14}$ Pa s is a linear approximation of the viscosity and $\tau_0$ is a user defined constant that ideally should be close to one.

We now turn to the implemented methods to impose the impenetrability condition.

### 3.1 Strong formulation

To enforce the impenetrability strongly, one would in the continuous case, if the boundary of the domain is sufficiently smooth, like the use test functions from a function space

$$\mathcal{V} = \left\{ \mathbf{v} \in [W^{1,\mathfrak{p}}(\Omega)]^d : \mathbf{v} \cdot \mathbf{n}|_{\Gamma_b} = \mathbf{0} \right\}.$$

This is the same as specifying a homogeneous Dirichlet boundary condition, in the direction of the normal. Consider the boundary integral in (7), which we decompose into normal and tangential parts as

$$\begin{aligned}\int_{\Gamma_b} \mathbf{n} \cdot (\mathbf{S}(\mathbf{u}) - p\mathbf{I}) \cdot (\mathbf{v} \cdot \mathbf{n})\mathbf{n} \, d\Gamma + \sum_{i=1}^{2} \int_{\Gamma_b} \mathbf{n} \cdot (\mathbf{S}(\mathbf{u}) - p\mathbf{I}) \cdot (\mathbf{v} \cdot \mathbf{t_i})\mathbf{t_i} \, d\Gamma \\ = - \sum_{i=1}^{2} \int_{\Gamma_b} \beta^2 (\mathbf{u} \cdot \mathbf{t_i}) \cdot (\mathbf{v} \cdot \mathbf{t_i}) \, d\Gamma \\ = - \int_{\Gamma_b} \beta^2 \mathbf{u} \cdot \mathbf{v} \, d\Gamma, \end{aligned} \tag{11}$$

where the first equality is due to the slip boundary condition (6) and the second is due to the orthonormality of $\mathbf{n}$ and $\mathbf{t}_i$ and $\mathbf{v} \in \mathcal{V}$.

If the domain is polygonal (e.g. a mesh), the discrete normal, $\mathbf{n}_h$, is only well defined on the interior of the facets belonging to the boundary, where it is constant on each facet. However, for edges or vertices on $\Gamma_b$, this is not the case. Typically, degrees of freedom (DOFs) for the velocity space lie on these locations. Furthermore, applying a Dirichlet condition to the velocity vector is done component-wise.

To extend the definition of $\mathbf{n}_h$ so that it is well defined for the DOFs lying on $\Gamma_b$, we have used the approach of a weighted average. That is, using the set of basis functions $\{\boldsymbol{\phi}^j\}_{j=1}^{N_\mathcal{V}}$ of $\mathcal{V}_h^k$, where $N_\mathcal{V}$ are the number of velocity DOFs, we assemble the discrete vector

$$\overline{\mathbf{n}}^j = \int_\Omega \boldsymbol{\phi}^j \cdot \mathbf{n}_h \, d\Omega \quad (j = 1, \dots, N_\mathcal{V}).$$

From this vector we select only the DOFs that lie on $\Gamma_b$ which are then normalized at each velocity node. Letting the number of nodes (each velocity node consisting of the $d$ number of velocity DOFs) be denoted by $n_\mathcal{V}$, the result is that $\overline{\mathbf{n}} = (\overline{\mathbf{n}}_1, \dots, \overline{\mathbf{n}}_{n_\mathcal{V}})^T$ contains, at each node $j$ that is on $\Gamma_b$, a unit normal $\overline{\mathbf{n}}^j = (\overline{n}_x, \overline{n}_y, \overline{n}_z)^T$, or $\overline{\mathbf{n}}^j = (\overline{n}_x, \overline{n}_y)^T$ in two dimensions, based on the average of the facets where the corresponding basis function is non-zero.

To enforce the Dirichlet boundary condition in the orientation of $\overline{\mathbf{n}}$, we proceed by a local coordinate transformation (John, 2002). We need to re-express the $d$ number of components of $\mathbf{u}_h$, given in the Cartesian coordinate system, to the at each node locally orthonormal coordinate system $(\overline{\mathbf{n}}, \overline{\mathbf{t}}_1, \overline{\mathbf{t}}_2)$, or $(\overline{\mathbf{n}}, \overline{\mathbf{t}}_1)$, where $\overline{\mathbf{t}}_i$ are the set of tangent vectors spanning the plane orthogonal to $\overline{\mathbf{n}}$. Since we are projecting $\mathbf{u}_h$ onto an orthogonal basis, the components in the new system become $(\mathbf{u}_h \cdot \overline{\mathbf{n}}, \mathbf{u}_h \cdot \overline{\mathbf{t}}_1, \mathbf{u}_h \cdot \overline{\mathbf{t}}_2)^T$, or $(\mathbf{u}_h \cdot \overline{\mathbf{n}}, \mathbf{u}_h \cdot \overline{\mathbf{t}}_1)^T$ in two dimensions. From this it can be seen that the desired Dirichlet boundary condition can be achieved by setting the first component to zero. As for the discrete tangent vectors, for $d = 2$ this is given as $\mathbf{t}_1 = (\overline{n}_z, -\overline{n}_x)^T$. For $d = 3$ we follow the algorithm presented in (John, 2002), which is shown in Algorithm 1 for completeness.

**Algorithm 1** Procedure to define, at each node, unit tangential vectors $\overline{\mathbf{t}}_1 = (\overline{t}_{1x}, \overline{t}_{1y}, \overline{t}_{1z})^T$ and $\overline{\mathbf{t}}_2 = (\overline{t}_{1x}, \overline{t}_{1y}, \overline{t}_{1z})^T$ orthonormal to $\overline{\mathbf{n}} = (\overline{n}_x, \overline{n}_y, \overline{n}_z)^T$

**if** $|\overline{n}_x|>0.5$ OR $|\overline{n}_y|>0.5$ **then**
    $n = \sqrt{\overline{n}_x^2 + \overline{n}_y^2}$
    $\overline{t}_{1x} = \overline{n}_y/n; \quad \overline{t}_{1y} = -\overline{n}_x/n; \quad \overline{t}_{1z} = 0$ ▷ (tangent $\overline{\mathbf{t}}_1$)
    $\overline{t}_{2x} = -\overline{t}_{1y}\overline{n}_z; \quad \overline{t}_{2y} = \overline{t}_{1x}\overline{n}_z; \quad \overline{t}_{2z} = \overline{t}_{1y}\overline{n}_x - \overline{t}_{1x}\overline{n}_y$ ▷ (tangent $\overline{\mathbf{t}}_2$)
**else**
    $n = \sqrt{\overline{n}_y^2 + \overline{n}_z^2}$
    $\overline{t}_{1x} = 0; \quad \overline{t}_{1y} = -\overline{n}_z/n; \quad \overline{t}_{1y} = \overline{n}_y/n$ ▷ (tangent $\overline{\mathbf{t}}_1$)
    $\overline{t}_{2x} = \overline{t}_{1z}\overline{n}_y - \overline{t}_{1y}\overline{n}_z; \quad \overline{t}_{2y} = -\overline{t}_{1z}\overline{n}_1; \quad \overline{t}_{2z} = \overline{t}_{1y}\overline{n}_1;$ ▷ (tangent $\overline{\mathbf{t}}_2$)
**end if**

A local transformation for node $j$ can therefore be given by the $d\times d$ matrix (we have here, just like in Algorithm 1 excluded the node indexing on the components, since this is clear from the context)

$$\mathbf{R}^j = \begin{cases} \begin{bmatrix} \overline{n}_x & \overline{n}_y & \overline{n}_z \\ \overline{t}_{1x} & \overline{t}_{1y} & \overline{t}_{1z} \\ \overline{t}_{2x} & \overline{t}_{2y} & \overline{t}_{2z} \end{bmatrix} & \text{if } d = 3 \text{ and } j \in \Gamma_b \\ \begin{bmatrix} \overline{n}_x & \overline{n}_z \\ \overline{t}_{1x} & \overline{t}_{1z} \end{bmatrix} & \text{if } d = 2 \text{ and } j \in \Gamma_b. \end{cases}$$

If the node $j$ does not lie on $\Gamma_b$, we simply set $\mathbf{R}^j = \mathbf{I}_{d\times d}$ to the identity matrix. The complete transformation matrix then becomes

$$\mathbf{R} = \begin{bmatrix} \mathbf{R}^1 & & \\ & \ddots & \\ & & \mathbf{R}_{n\mathcal{V}} \end{bmatrix},$$

that due to its construction has the property $\mathbf{R}\mathbf{R}^T = \mathbf{I}$.

In matrix form, (8) becomes

$$\begin{bmatrix} \mathbf{C} & \mathbf{B}^T \\ \mathbf{B} & \mathbf{0} \end{bmatrix} \begin{bmatrix} \mathbf{u}_h \\ \mathbf{p}_h \end{bmatrix} = \begin{bmatrix} \mathbf{f} \\ \mathbf{0} \end{bmatrix}, \tag{12}$$

where $\mathbf{B}$ is the discrete divergence operator and $\mathbf{C}$ is the (linearized) stiffness matrix with the addition of the boundary integral (11). If we number $\mathbf{u}_h$ as

$$\mathbf{u}_h = (\mathbf{u}_h^1, \ldots, \mathbf{u}_h^{n\mathcal{V}})^T,$$

where each $\mathbf{u}_h^j = (u_h^x, u_h^y, u_h^z)$ is a vector of components of $\mathbf{u}_h$ at the $j$:th node, we can apply the transformation as

$$\begin{bmatrix} \mathbf{R} & \mathbf{0} \\ \mathbf{0} & \mathbf{I} \end{bmatrix} \begin{bmatrix} \mathbf{C} & \mathbf{B}^T \\ \mathbf{B} & \mathbf{0} \end{bmatrix} \begin{bmatrix} \mathbf{R} & \mathbf{0} \\ \mathbf{0} & \mathbf{I} \end{bmatrix}^T \begin{bmatrix} \mathbf{R} & \mathbf{0} \\ \mathbf{0} & \mathbf{I} \end{bmatrix} \begin{bmatrix} \mathbf{u}_h \\ \mathbf{p}_h \end{bmatrix} = \begin{bmatrix} \mathbf{R} & \mathbf{0} \\ \mathbf{0} & \mathbf{I} \end{bmatrix} \begin{bmatrix} \mathbf{b}_h \\ \mathbf{0} \end{bmatrix}.$$

This can be reduced to

$$\begin{bmatrix} \overline{\mathbf{C}} & \overline{\mathbf{B}}^T \\ \overline{\mathbf{B}} & \mathbf{0} \end{bmatrix} \begin{bmatrix} \overline{\mathbf{u}}_h \\ \mathbf{p}_h \end{bmatrix} = \begin{bmatrix} \overline{\mathbf{b}}_h \\ \mathbf{0} \end{bmatrix},$$

where $\overline{\mathbf{C}} = \mathbf{R}\mathbf{C}\mathbf{R}^T$, $\overline{\mathbf{B}} = \mathbf{B}\mathbf{T}^T$, $\overline{\mathbf{b}}_\mathbf{h} = \mathbf{R}\mathbf{b}_h$ and $\overline{\mathbf{u}}_\mathbf{h} = \mathbf{R}\mathbf{u}_h$. After this system is solved, one can retrieve the original solution by $\mathbf{u}_h = \mathbf{R}^T\overline{\mathbf{u}}_h$.

### 3.2 Weak formulation

The main idea behind imposing impenetrability weakly, is to use a Lagrange multiplier in a similar way to how the pressure is used to impose incompressibility. However, contrary to the pressure which is defined on the whole domain, the Lagrange multiplier enforcing the impenetrability should only be defined on the slip boundary, $\Gamma_b$.

The role of the Lagrangian multipliers in the context of (1), comes from considering the problem as that of finding a saddle-point of the Lagrangian functional. For the case of ice as a free surface flow, the functional $\mathcal{L}(\mathbf{u}, p)$ has been determined in e.g. Dukowicz *et al.* (2010), in which it is shown that this process is equivalent to minimizing the energy of the system. Further, in Dukowicz *et al.* (2011), the additional constraint of impenetrability is incorporated in the functional. By introducing the function space $\Lambda := W^{\frac{1}{\mathrm{p}^*},\mathrm{p}}(\Gamma_b)$, consisting of the traces of functions in $\mathcal{V}$, the problem becomes to seek a solution $\mathbf{u} \in \mathcal{V}$, $p \in \mathcal{Q}$ and $\lambda \in \Lambda$ such that

$$\mathcal{L}(\mathbf{u}, p, \lambda) := \inf_{\substack{q \in \mathcal{Q} \\ \varrho \in \Lambda}} \sup_{\mathbf{v} \in \mathcal{V}} \mathcal{L}(\mathbf{v}, q, \varrho). \tag{13}$$

We will not derive the functional in its entirety (it is derived in Dukowicz *et al.* (2010, 2011)), but simply state that the above amounts to setting the functional (Gâteaux) derivative of $\mathcal{L}(\mathbf{u}, p, \lambda)$ to zero as

$$\begin{aligned}
&\lim_{\varepsilon \to 0} \frac{d}{d\varepsilon} \mathcal{L}(\mathbf{u} + \varepsilon \mathbf{v}, p + \varepsilon q, \lambda + \varepsilon \varrho) = \\
&\lim_{\varepsilon \to 0} \frac{d}{d\varepsilon} \left[ \int_\Omega \frac{2n}{n+1} \eta(\mathbf{II}_\mathrm{D}) \mathbf{II}_\mathrm{D} - (p + \varepsilon q) \nabla \cdot (\mathbf{u} + \varepsilon \mathbf{v}) - \rho \mathbf{g} \cdot (\mathbf{u} + \varepsilon \mathbf{v}) \, d\Omega \right. \\
&\qquad \left. + \int_{\Gamma_b} (\lambda + \varepsilon \varrho)(\mathbf{u} + \varepsilon \mathbf{v}) \cdot \mathbf{n} + \beta^2 (\mathbf{u} + \varepsilon \mathbf{v}) \, d\Gamma_b \right] \\
&\qquad = 0 \quad \forall (\mathbf{v}, q, \varrho) \in \mathcal{V} \times \mathcal{Q} \times \Lambda,
\end{aligned}$$

where it is understood that $\mathbf{II}_\mathrm{D}$ depends on $\mathbf{u} + \varepsilon \mathbf{v}$. The above results in (1a), with the additional constraints

$$-\int_\Omega q \nabla \cdot \mathbf{u} \, d\Omega = 0 \quad \forall q \in \mathcal{Q}, \tag{14a}$$

$$\int_{\Gamma_b} \varrho \mathbf{u} \cdot \mathbf{n} \, d\Gamma_b = 0 \quad \forall \varrho \in \Lambda, \tag{14b}$$

$$\int_{\Gamma_b} \Big( \mathbf{n} \cdot (\mathbf{S}(\mathbf{u}) - p\mathbf{I}) + \beta^2 \mathbf{u} + \lambda \mathbf{n} \Big) \cdot \mathbf{v} \, d\Gamma_b = 0 \quad \forall \mathbf{v} \in \mathcal{V}. \tag{14c}$$

The integrals (14a) and (14b) above give the incompressibility condition and that $\mathbf{u} \cdot \mathbf{n} = 0$ on $\Gamma_b$, as desired. Similarly we must have that

$$\mathbf{n} \cdot (\mathbf{S}(\mathbf{u}) - p\mathbf{I}) + \beta^2 \mathbf{u} + \lambda \mathbf{n} = 0.$$

Multiplying the above by the unit normal eliminates the $\beta^2$-term (due to the impenetrability) giving

$$\lambda = -\mathbf{n} \cdot (\mathbf{S}(\mathbf{u}) - p\mathbf{I}) \cdot \mathbf{n} = -\mathbf{n} \cdot \mathbf{T} \cdot \mathbf{n}.$$

From this we see that the value of $\lambda$ has the physical meaning of the negative normal component of the Cauchy stress tensor. Hence, adding the Lagrange multiplier term is equivalent to subtracting the normal component of the Cauchy stress tensor, making the remainder relating the tangential terms of $\mathbf{T}$ to the slip velocity. Using both surface integrals in (14) to replace the surface integral in (7), results in the weak formulation

$$\begin{aligned}\int_\Omega \mathbf{S}(\mathbf{u}) : \nabla \mathbf{v}\, d\Omega - \int_\Omega p \nabla \cdot \mathbf{v}\, d\Omega \\ + \int_{\Gamma_b} \beta^2 \mathbf{u} \cdot \mathbf{v}\, d\Gamma_b \\ + \int_{\Gamma_b} \lambda \mathbf{v} \cdot \mathbf{n}\, d\Gamma_b = \int_\Omega \rho \mathbf{g} \cdot \mathbf{v}\, d\Omega, \quad \forall \mathbf{v} \in \mathcal{V}\end{aligned} \tag{15a}$$

$$\int_\Omega \nabla \cdot \mathbf{u} q\, d\Omega = 0, \quad \forall q \in \mathcal{Q} \tag{15b}$$

$$\int_{\Gamma_b} \varrho \mathbf{u} \cdot \mathbf{n}\, d\Gamma_b = 0 \quad \forall \varrho \in \Lambda_h. \tag{15c}$$

In this study we have for the space approximating $\Lambda$ chosen the space of polynomials that are piecewise constant on the facets of $\Omega$. That is,

$$\Lambda_h := \left\{ \varrho \in L^2(\Gamma_b) : \varrho|_F \in \mathcal{P}_0, F \in \mathcal{F}_h \right\}.$$

At this stage, a few notes regarding the above mentioned method should be recognized. First, since a Lagrange multiplier is used to enforce the impenetrability condition, we have added an *additional unknown*. In the discrete case, this increases the number of DOFs resulting in a larger and more expensive system to solve. Second, as can be seen from (13), the saddle-point character of the system is enhanced. This has for the linear Stokes problem been studied in Verfürth (1986), where the the author shows that the above problem may suffer from instabilities in the Lagrange multiplier parameters ($p$ and $\lambda$) even for choices of finite element families that satisfy the traditional *inf-sup* condition for (7). The author shows that this can be remedied by either enriching the velocity space (Verfürth, 1986) or stabilizing the formulation by adding a GLS term to the relevant facets lying on $\Gamma_b$ (Verfürth, 1991). The effect of the stabilized method has been studied in (Urquiza *et al.*, 2014). The latter study suggests that, on domains with polygonal approximations of curved boundaries, convergence towards an exact solution is dependent of the choices of the stabilization parameters.

Unfortunately, we were not able to implement either method suggested above. We have dealt with this issue by examining if either $p$ or $\lambda$ have shown any behavior resembling instability in the simulations. This is an obvious shortcoming of the present study. In the two-dimensional simulations, this did not become a problem. However, in three dimensions, it seems that stabilization is necessary. This can possibly be related to that the $P2P0$ formulation is stable in $d = 2$, but not in $d = 3$.

### 3.3 Approximate formulation

Below we present the final method, which has been used in Brinkerhoff and Johnson (2013). The method essentially starts from the Lagrangian functional presented in Section 3.2, but instead of introducing a new variable as a Lagrange multiplier to enforce the impenetrability at the bed, it uses the pressure variable to do this. The advantages of this method is in that neither a local transformation (Section 3.1) nor increasing the number of unknowns of the system (Section 3.2) is necessary, and that the implementation is essentially inherent in every FEM software. The drawback is its approximative character, which can possibly affect both the pressure and velocity field.

Consider the Lagrangian functional in (14), but as a function of only $\mathbf{v}$ and $p$ as

$$\mathcal{L}(\mathbf{u}, p, p) := \inf_{q \in \mathcal{Q}} \sup_{\mathbf{v} \in \mathcal{V}} \mathcal{L}(\mathbf{v}, q, q). \tag{16}$$

Setting the Gâteaux derivative equal to zero again results in (1). However, since only the pressure variable is used, the analogues to (14a) and (14b) is to combine these into one condition, while (14c) in this case becomes independent of $p$, as

$$-\int_\Omega q \nabla \cdot \mathbf{u} \, d\Omega + \int_{\Gamma_b} q \mathbf{u} \cdot \mathbf{n} \, d\Gamma_b = 0 \quad \forall q \in \mathcal{Q}, \tag{17a}$$

$$\int_{\Gamma_b} \Big( \mathbf{n} \cdot (\mathbf{S}(\mathbf{u}) - p\mathbf{I}) + \beta^2 \mathbf{u} + p\mathbf{n} \Big) \cdot \mathbf{v} \, d\Gamma_b =$$
$$\int_{\Gamma_b} \Big( \mathbf{n} \cdot \mathbf{S}(\mathbf{u}) + \beta^2 \mathbf{u} \Big) \cdot \mathbf{v} \, d\Gamma_b = 0 \quad \forall \mathbf{v} \in \mathcal{V}. \tag{17b}$$

Since we only have Neumann type boundary conditions (i.e. we have not specified the velocity or pressure anywhere on the boundary), the above equations give rise to the following compatibility condition

$$\int_\Omega \nabla \cdot \mathbf{u} \, d\Omega = \int_{\Gamma_b} \mathbf{u} \cdot \mathbf{n} \, d\Gamma_b \quad \stackrel{\text{IBP}}{\Longrightarrow}$$
$$-\int_\Omega \nabla 1 \cdot \mathbf{u} \, d\Omega + \int_\Gamma 1 \mathbf{u} \cdot \mathbf{n} \, d\Gamma = \int_{\Gamma_b} \mathbf{u} \cdot \mathbf{n} \, d\Gamma_b \quad \Rightarrow$$
$$\int_{\Gamma_s} \mathbf{u} \cdot \mathbf{n} \, d\Gamma_s = 0 \tag{18}$$

This can be seen as relaxing the normal condition for conservation of mass. Thus, impenetrability is not necessarily strictly enforced and if the normal velocity is non-zero this affects the divergence of the velocity field in the area directly connected to $\Gamma_b$. By (17b) we can see that the added term eliminates the pressure variable, thus relating the deviatoric stress at the boundary to the *total* velocity. That is, the above does not result in the boundary condition in (6), since this only relates the tangential components of $\mathbf{S}$ to the tangential velocity. The weak formulation is naturally obtained by replacing $(\lambda, \varrho)$ with $(p, q)$ in (15).

# 4 Results

We start in two dimensions by using the method of manufactured solutions presented in Sargent and Fastook (2010) (and subsequently extended in Leng *et al.* (2013)), which indicates similar convergence rates and solutions of all the methods presented in Section 3 to the analytical solution. Following this, we consider the sliding benchmark scenario of *Haute Glacier d'Arolla* (Pattyn *et al.*, 2008). Contrary to the convergence studies, this simulation shows a different solution for the approximative method compared to the other two methods, which is more in line with the analysis in Section 3.3. Finally we use the strong and approximative method to simulate the Greenland Ice Sheet for a given $\beta^2$-field and compare the two solutions. In Table 1 values for the different simulations performed herein highlighting and confirming the analysis made in the previous section are presented.

All simulations in this study where performed with the FEM software package FEniCS (Logg *et al.*, 2012; Alnæs *et al.*, 2015).

Table 1. The measures of incompressibility and impenetrability for the strong (**S**), weak (**W**) and approximative (**A**) methods for each of the simulated scenarios: manufactured solutions (**Man**), Glacier Haute d'Arolla (**HdA**) and the Greenland Ice Sheet **GrIS**.

| | $\int_\Omega \nabla \cdot \mathbf{u}$ | | | $\int_{\Gamma_b} \mathbf{u} \cdot \mathbf{n}$ | | |
|---|---|---|---|---|---|---|
| | **S** | **W** | **A** | **S** | **W** | **A** |
| **Man** | 1.42e−11 | 1.92e−11 | −2.68e−2 | 1.22e−14 | −7.93e−14 | −2.68e−2 |
| **HdA** | −2.00e−12 | −2.75e−12 | 1.38e1 | −4.31e−14 | −5.55e−13 | 1.38e1 |
| **GrIS** | 7.67e3 | | 2.22e9 | −7.87e2 | | 2.22e9 |

### 4.1 Convergence rates for the manufactured solutions

To examine to convergence rates of the methods, we use the two-dimensional *steady-state* manufactured solutions from Sargent and Fastook (2010), given for the velocity $\mathbf{u} = (u, w)$ and pressure $p$. For a given geometry described by a surface, $z_s(x)$, and a bed, $z_b(x)$, the solutions are given by

$$\begin{aligned} u(x,z) &= \frac{c_x Z}{z_s - z_b}\left[1 - \left(\frac{z_s - z}{z_s - z_b}\right)^{\lambda_1}\right] + c_b \frac{1}{z_s - z_b} \\ w(x,z) &= u(x,z)\left[\frac{dz_b}{dx}\frac{z_s - z}{z_s - z_b} + \frac{dz_s}{dx}\frac{z - z_b}{z_s - z_b}\right] \\ p(x,z) &= -2\eta\frac{\partial u}{\partial x} + \rho g (z_s - z), \end{aligned} \tag{19}$$

where $c_x$ and $c_b$ are parameters that specify the velocity due to deformation at the surface and scaled sliding speed and $\eta$ is calculated by (3).

For the scenario we choose an ISMIP-HOM type domain (Pattyn *et al.*, 2008) as

$$\begin{aligned} z_s(x) &= -\tan(\alpha) \\ z_b(x) &= z_s(x) - Z + \frac{Z}{4}\sin\left(\frac{2\pi x}{L}\right), \end{aligned}$$

with $L = 8000\,\mathrm{m}$ and $Z = 400\,\mathrm{m}$ being the length and typical height of the domain and $\alpha = 1°$. For the considered simulation we specified $c_x = 40$ and $c_b = 30$ (corresponding to a typical surface velocity of $70\,\mathrm{m/a}$ ) and chose the basal drag coefficient as

$$\beta^2 = 200 + 1000\left(1 - \sin\left(\frac{2\pi x}{L}\right)\right)$$

in an attempt to minimize the introduced basal stresses added by the manufactured solutions. The manufactured solutions by design fulfill the impenetrability condition (5), however the appropriate body force and boundary conditions need to be determined by inserting (19) into (1), (4) and (6) respectively. We have here followed the method presented in Leng *et al.* (2013) and related supplementary material, in which we set a maximum values for the viscosity to be $\eta_{\max} = 10^{10}\,\mathrm{Pa\,a}$.

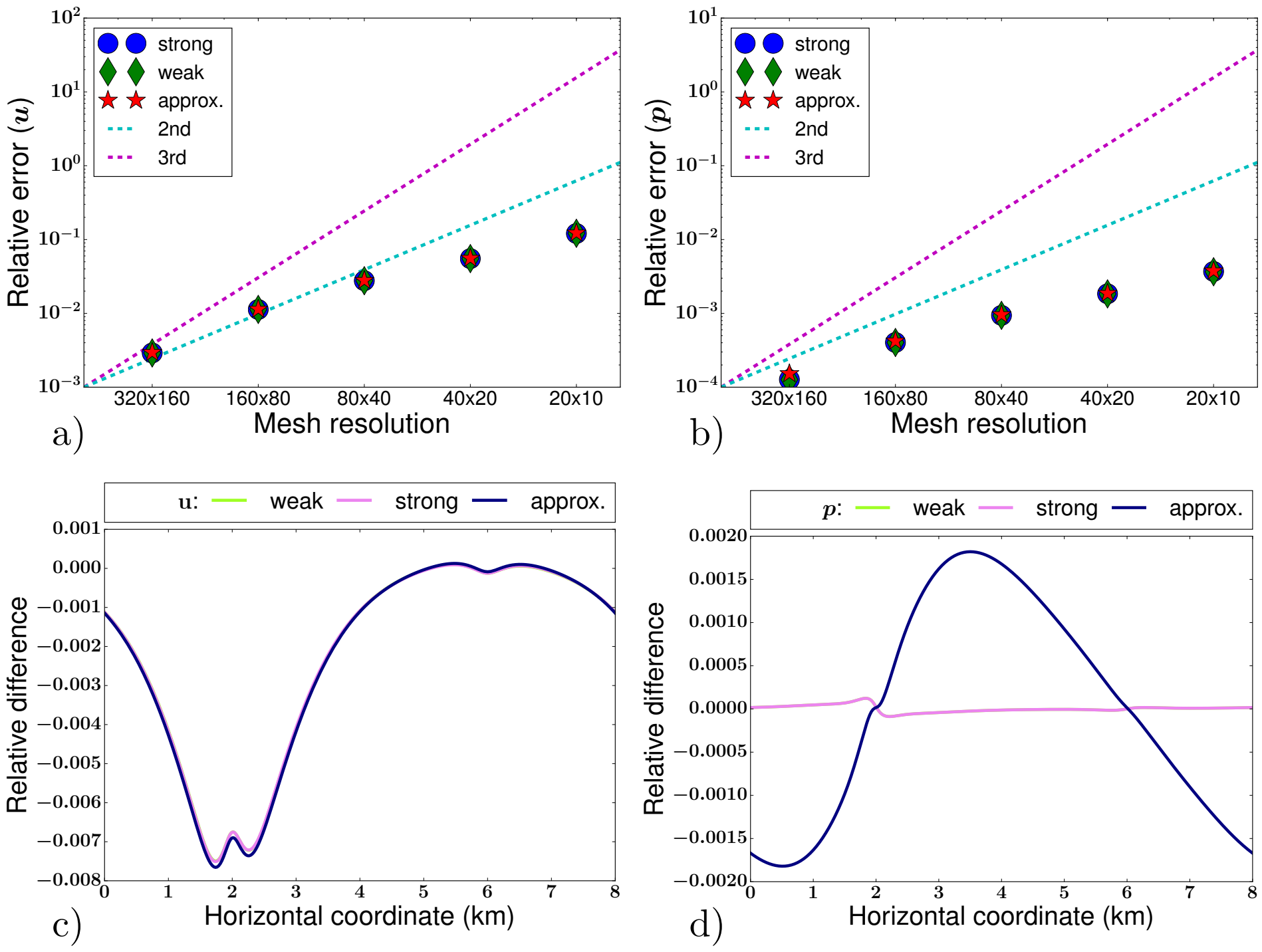

Figure 1. Comparison between the simulated and manufactured solutions using the $P2P1$ element. Top panel (a and b) shows the convergence rates for velocity and pressure using the relative error in the $L^2$-norm. Dashed lines show theoretical 2nd and 3rd order convergence rates. Bottom panel (c and d) shows relative difference between the manufactured solution and the simulated basal velocity (c) and pressure at the bed (d) for each method. While the approximative methods shows slightly different results compared to the strong and weak methods (almost indiscernible in the figure), the differences are minor and do not affect the convergence.

Figure 1a and Fig. 1b shows the convergence rates for the scenario introduced above. We use the stable $P2P1$ formulation which for the linear Stokes problem theoretically should give 2nd order convergence rate for the *gradient* of the velocity and pressure, but can be extended to 3rd order for the velocity in the case of convex polygons in $d = 2$ (e.g. Ern and Guermond, 2004). For the non-linear problem the convergence rates should be measured in the norms of $\mathcal{W}^{1,\mathfrak{p}}$ and $\mathcal{L}^{\mathfrak{p}^*}$ (e.g. Hirn, 2011), but we choose to follow the glaciological literature and to present the convergence rates in the $L^2$-norm connected to the linear problem (e.g. Brinkerhoff and Johnson, 2013; Leng *et al.*, 2013; Gagliardini *et al.*, 2013). As can be seen, although all methods converge towards the manufactured solution in a practically identical manner, the convergence rate for the velocity is suboptimal. The reason for this could possibly be that the manufactured solutions contain singularities at the surface at the $x$-coordinates $L/4$ and $3L/4$. This results in (theoretically) infinite values in the compensation terms (introduced surface traction). Clearly this is not desirable in a numerical implementation and, even though the magnitude of the compensation terms is limited by $\eta_{\max}$, we do not exclude that this is the reason for the poor convergence rate for the velocity. The relative differences between the solutions of each method and the manufactured solution are very similar as well (Fig. 1c and Fig. 1d). Even though the approximative method differs from the strong and weak method, particularly for the pressure, the differences are of small magnitude (for a comparison of the errors of the velocity fields, see Fig. A1 in Appendix A). For the no-slip scenario ($\mathbf{u}_b = 0$), convergence rates found in Leng *et al.* (2013) for a similar but three-dimensional case agreed well with the theoretical rates. However, in Gagliardini *et al.* (2013), convergence rates for the same setup with first order bi-linear elements ($Q1Q1$) was found to be of 3rd order, that is *higher* than what is theoretically expected.

We also note that we have done the same convergence study for the no-slip case and used both $P2P1$ and $P1P1$ elements (not shown). The no-slip case did not significantly change the convergence rates, but for the convergence rates for the $P1P1$ elements were close to the expected (2nd order for both velocity and pressure). Possibly, this indicates that the added compensation terms, specially around the singularity, changes the problem to a degree where either the solution is not regular enough to benefit from the extra order of convergence for the velocity or that the verification of the method using the manufactured solutions is not suitable.

### 4.2 A test case scenario: Haute Glacier d'Arolla

We follow the specifications presented in Pattyn *et al.* (2008) for the case of Haute Glacier d'Arolla with a slip zone. The basal drag coefficient is defined as as

$$\beta^2 = \begin{cases} 0 & \text{if } 2200 \le x \le 2500 \\ 10^{16} & \text{otherwise.} \end{cases}$$

Note that the high value of $\beta^2$ effectively sets the basal velocity $\mathbf{u}_b = \mathbf{0}$, even though we have not implemented this strictly as a Dirichlet boundary condition. We again used the $P2P1$ element while the mesh used was an unstructured triangular mesh with a characteristic cell size of 10 m generated by Gmsh (Geuzaine and Remacle, 2009). This should be considered as a fairly fine mesh (contrary to the general specifications given in Pattyn *et al.* (2008), which are $\Delta x = 100$ m). The reason for this is to emphasize the *difference* between the methods on Glacier Haute d'Arolla compared to the *similarities* indicated by the convergence studies presented in Section 4.1. In Fig. 2 the basal velocity and basal pressure over the slip zone are presented. In the case of the fine mesh shown, the strong and weak methods are practically identical (indiscernible from each other in the figure), while the approximative method differs over the slip region. Given the presentation in Section 3 it is expected that the strong and weak methods should converge to the same result. The approximative solution shows a slightly lower basal velocity than the other methods (Fig. 2a), while from Fig. 2b it becomes clear that the more significant difference (at least in relative terms) lies in the pressure. This is also supported by the conditions (17a) and (17b), that indicate that the pressure should change when sliding is present.

Due to the similarity of the solutions of the strong and weak methods, we next focus on the difference between the strong and approximative methods. This is shown in Fig. 3, that depicts the norm of the differences of the velocity and pressure fields. In essence, the above argument is further elucidated by this. The relative difference of the velocity field in the slip zone is in the order of 0.01 while it is slightly higher, around 0.05 for the pressure field. Note that the pressure field is more or less identical outside the slip zone and away from the bed while the velocity field is affected both up- and down-stream. If compared to the inter comparison results presented in Pattyn *et al.* (2008), all of the methods examined in this study fall within the range of the FS results. It therefore becomes

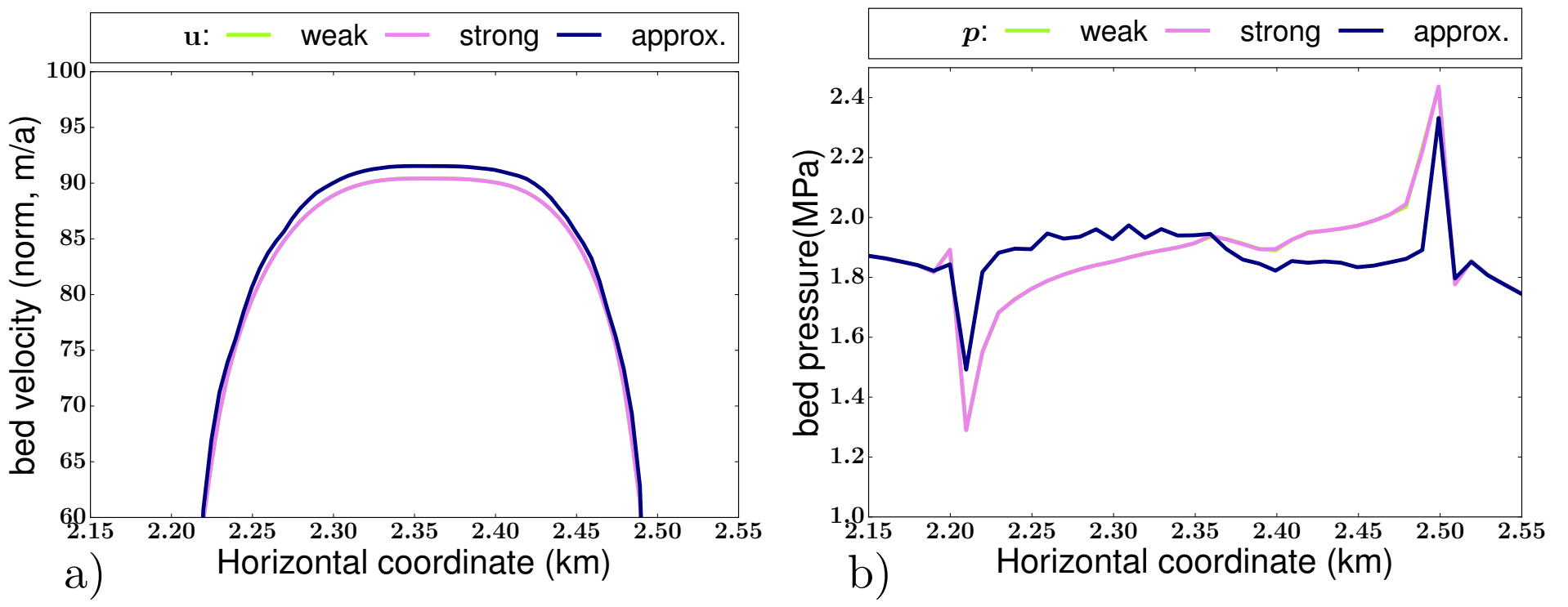

Figure 2. Haute Glacier d'Arolla simulations, basal velocity and pressure for all three methods (strong, weak and approximative). Figures are focused on the slip zone $2200 \leq x \leq 2500$. Panel (a) shows the norm of the basal velocity and panel (b) shows the basal pressure. In both panels the strong and weak methods nearly coincide

somewhat difficult to justify one method over the other solely based on the benchmark glacier simulation. We also note that this scenario is somewhat extreme, representing a glacier frozen to its bed except in a limited zone where free slip occurs. To evaluate the methods on a more standard scenario, we next model the Greenland Ice Sheet (GrIS) for a specified $\beta^2$-field.

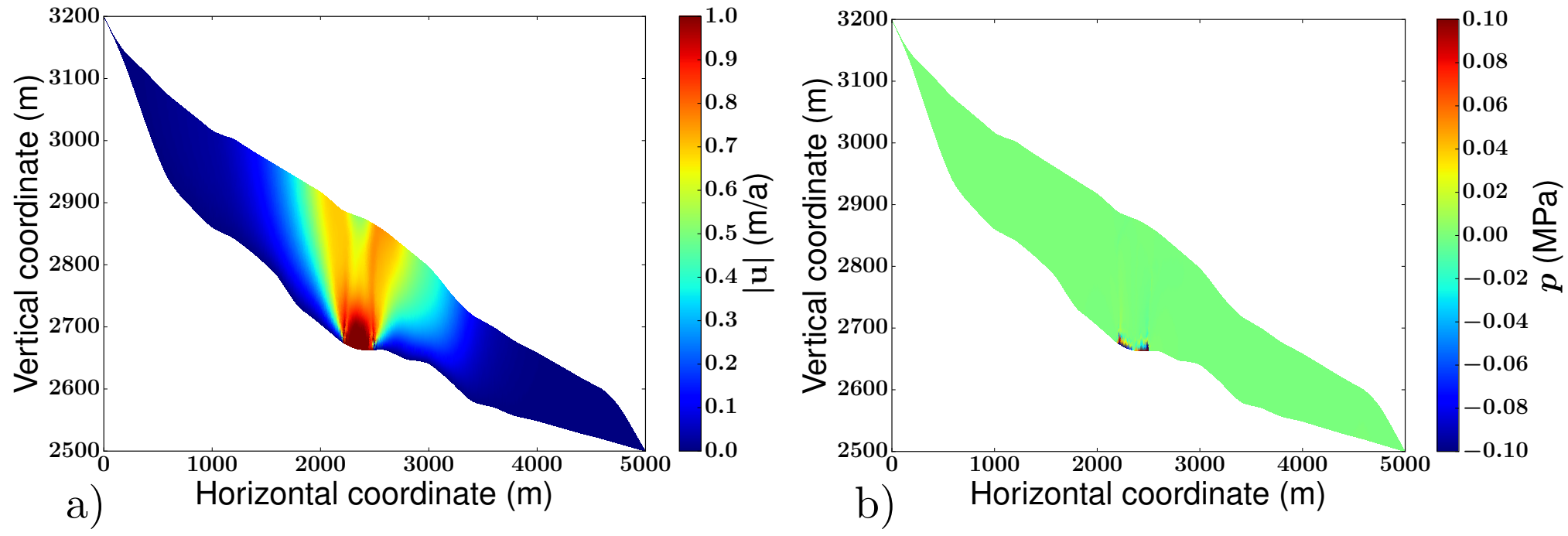

Figure 3. Differences between the strong and approximative method for the Haute Glacier d'Arolla case. The panels show the norm of the velocity difference (a) and pressure difference (b).

### 4.3 A realistic ice sheet simulation: Greenland

For simulating the GrIS we used a mesh consisting of 414 585 elements (tetrahedrons), based on an unstructured triangular partition in the horizontal domain (9213 triangles) extruded to 15 layers in the vertical. This mesh was subsequently deformed to the surface and bed topography of the GrIS. The footprint mesh and the $\beta^2$-field are shown in Fig. 4b. Both the footprint mesh and $\beta^2$-field were supplied by J. Ahlkrona (personal communication).

Due to limited access to computational power and the relatively high computational costs that it takes to perform a three-dimensional simulation, we have for the GrIS used linear elements for both the velocity and pressure ($P1P1$). As remarked in Section 3, contrary to the $P2P1$ formulation used for (most) of the two-dimensional simulations, the $P1P1$ formulation needs to be stabilized. In the GrIS simulations we have therefore added the terms in (9) to (7), and in (10) chosen to define $\tau_0 = 1$.

The simulations were performed with the strong and approximative method. The reasons for this are as follows: the stability issues the weak method is prone to (see Section 3.2) were affecting the output in a dominant way. Unfortunately, this issue had not been resolved at the time of writing. Furthermore, both the added computational costs and the similarity of the strong and weak methods shown in Fig. 1, served as an argument that performing simulations for both the strong and weak method was not necessary. Lastly, to the authors' knowledge, it is the strong and approximative methods that have been used in the field of numerical glaciology (e.g. Gagliardini *et al.* (2013); Leng *et al.* (2012) for strong and Brinkerhoff and Johnson (2013) for approximative).

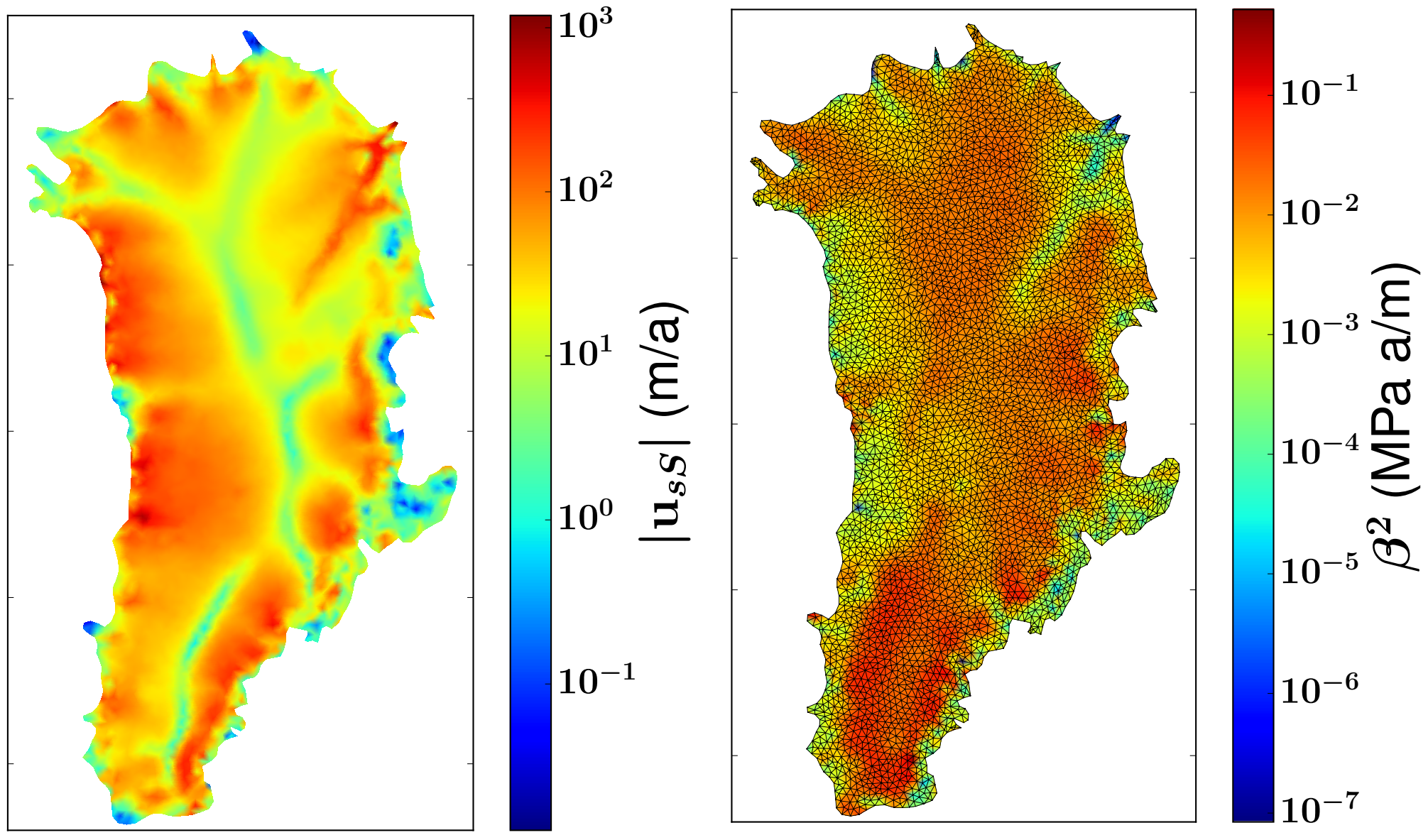

a) b)

Figure 4. Output and mesh used for the GrIS simulations. Panel (a) show the norm of the surface velocity simulated using the strong method. Panel (b) shows the $\beta^2$-field and footprint mesh used for the simulations.

In the figures presented, we use the subscripts $s$ and $b$ to denote velocity or pressure at the surface and bed, and will similarly use subscripts $S$ and $A$ to specify that the output is simulated using the strong or approximative method. In Fig. 4a the norm of the surface velocity field simulated using the strong method is shown ($\mathbf{u}_{sS}$). This figure shows the expected general pattern of faster flow towards the coastal regions, especially where marine terminating outlet glaciers are located. Since we do not have access to a true solution of the problem, we consider the output of the strong method as a reference for the comparison to the solution resulting from the approximative method. As a measure, we define the *relative difference* of a variable as

$$RD(\mathbf{u}) = \frac{|\mathbf{u}_S - \mathbf{u}_A|}{|\mathbf{u}_S|},$$

with $|\cdot|$ being the Euclidean norm. The relative differences for the basal (sliding) velocity ($\mathbf{u}_b$) and basal pressure ($p_b$) are shown in Fig. 5. The differences for the velocity and pressure fields show similar patterns, as would be expected. The high differences in the velocity field do not strictly follow the areas of fast flow. Naturally, in areas where flow is very low, the relative measure $RD(\mathbf{u}_b)$ is prone to becoming large when absolute differences in velocity are very small. However, in areas of moderate to high flow (greenish to yellowish in Fig. 5a), $RD(\mathbf{u}_b) \approx 10^{-2}$–$10^{-1}$ which has a notable effect on the resulting surface velocities in these areas. The oscillations in basal pressure present in the output of the approximative method are similarly most dominant in these areas. The highest values occur along the south east coastal regions. Since the $\beta^2$-values are fairly high (and as a result, velocities moderate), this is probably due to the varying bottom topography present (in particular for this resolution of the mesh). For completeness, we include a figure of the norm of differences for the basal velocities and pressures in Fig. A2 in Appendix A.

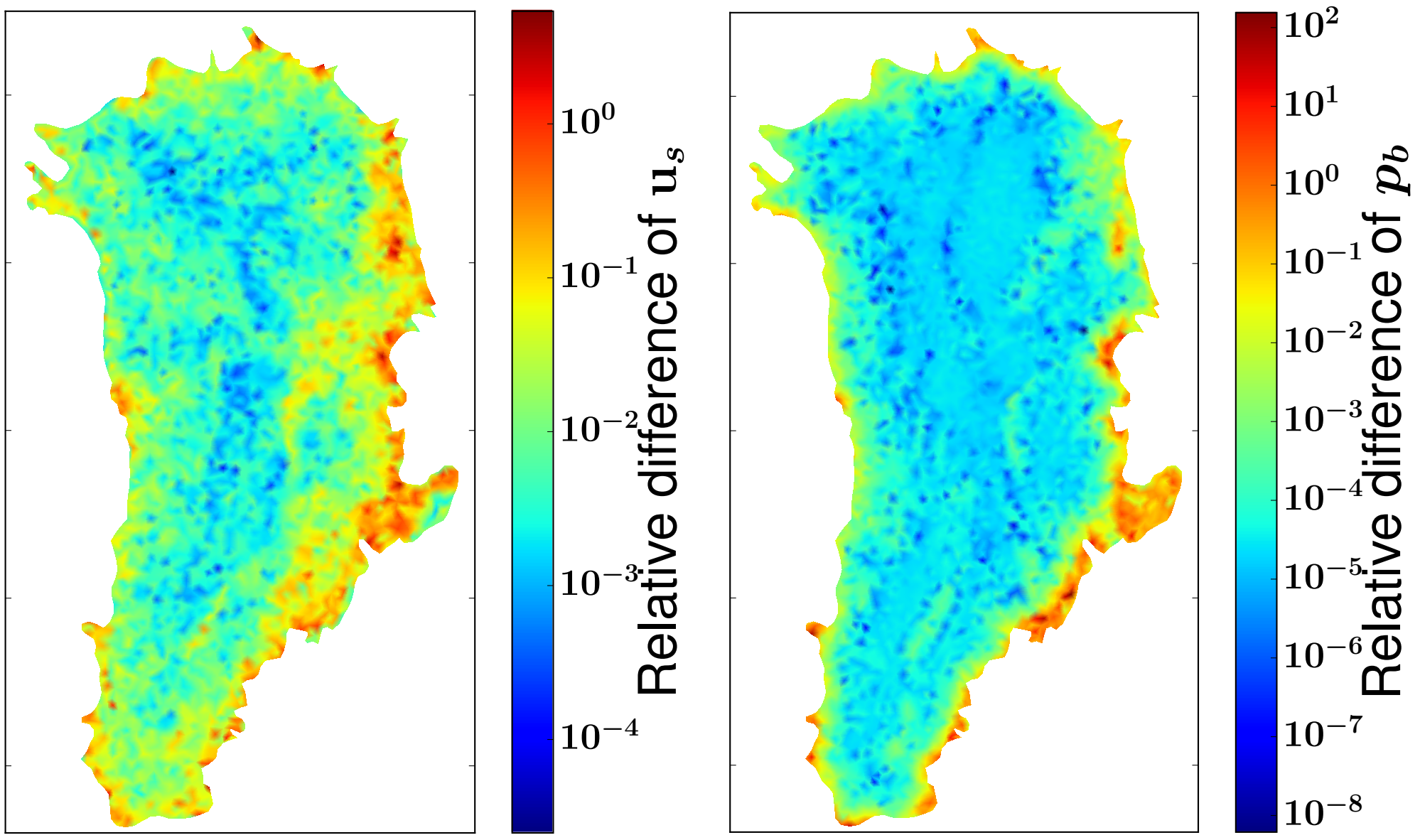

a) b)

Figure 5. The relative difference between the strong and approximative method for the basal velocity (a) and basal pressure (b).

## 5 Synthesis and Conclusions

Motivated by the importance of sliding at the bed of glaciers and ice sheets, we have in this study examined the output of three methods applied to the Full Stokes equations: the strong, weak and approximative method. The strong method has been used frequently for this purpose in the ice sheet community, while the approximative method is less common and the weak method is only known to have been previously used for a simple scenario in Dukowicz (2012). Using the same software has allowed us to directly compare the methods to one another.

We evaluate the methods by simulating Haute Glacier d'Arolla and the Greenland Ice Sheet. Given that the approximative method is very simple to implement the performance is acceptable when considering the differences compared to the more accurate strong method. In areas of significant slip or highly varying bed topography, the differences in velocity and pressure typically are in the order of $\sim 1\%$, but can rise to $> 5\%$, while in parts where flow and topography are moderate the approximative method is practically identical to the strong method. However, due to the nature of the method, the pressure variable along the sliding boundary will always be affected to some degree and therefore also affect the incompressibility by a small amount. The potential oscillations introduced at the bed when compared to the strong method in the considered GrIS scenario, are of low enough magnitude ($\pm 0.24\,\mathrm{MPa}$) to not affect any secondary variables, such as the pressure melting point or temperature at the bed. For the two-dimensional simulations, the strong an weak method are nearly identical for a fine mesh. In this study we have not evaluated the weak method in a three-dimensional setting and can therefore not exclude that this would differ from the strong method in e.g. the GrIS scenario, although we acknowledge that this is a possibility in particular for coarser meshes.

We have also used the method of manufactured solutions, to verify (the match or mismatch) of the methods. However, the similarity of the output of the different methods, most probably due to the large compensation terms added to the FS equations, we can only conclude that the use of this control method in the present scope seems to not be valid.

From a practical point of view, considering that the $\beta^2$-field used as part of the slip boundary condition most often is the result of an inversion method attempting to minimize the mismatch between some observed velocity data and the simulated surface velocity, the choice of any of the herein considered methods may be based on ease of implementation. However, if studying specific processes, the approximative method should be used with care.

# 6 Acknowledgments

Christian Helanow was supported by the nuclear waste management organizations in Sweden (Svensk Kärnbränslehantering AB) and Finland (Posiva Oy) through the Greenland Analogue Project and by Gålöstiftelsen. Peter Jansson and Stefan Bjursäter are thanked for their comments on the manuscript.

# Appendix A

Figure A1 shows the the absolute error in velocity of the strong, weak and approximative methods compared to the manufactured solutions for the ISMIP-HOM type simulation. The output of the different methods are very similar over the domain. This emphasizes the difference between using the manufactured solutions compared to the more realistic scenario of Haute Glacier d'Arolla. Since the norm is taken over the domain, it is likely that the small differences between the methods are not reflected in the convergence rates, giving an impression of the methods being similar. Figure A2 shows the absolute error in velocity, the difference $p_{bS} - p_{bA}$ and the pressure variable for the strong method for the GrIS simulation. Here we can see the areas by the coast differ by $10\,\mathrm{m\,a^{-1}}$–$100\,\mathrm{m\,a^{-1}}$. The pressure differences are small in the majority of the ice sheet domain, but relatively high in areas where the bed topography varies and sliding is significant. This is due to the pressure in the approximative method has the double role of enforcing incompressibility and impenetrability.

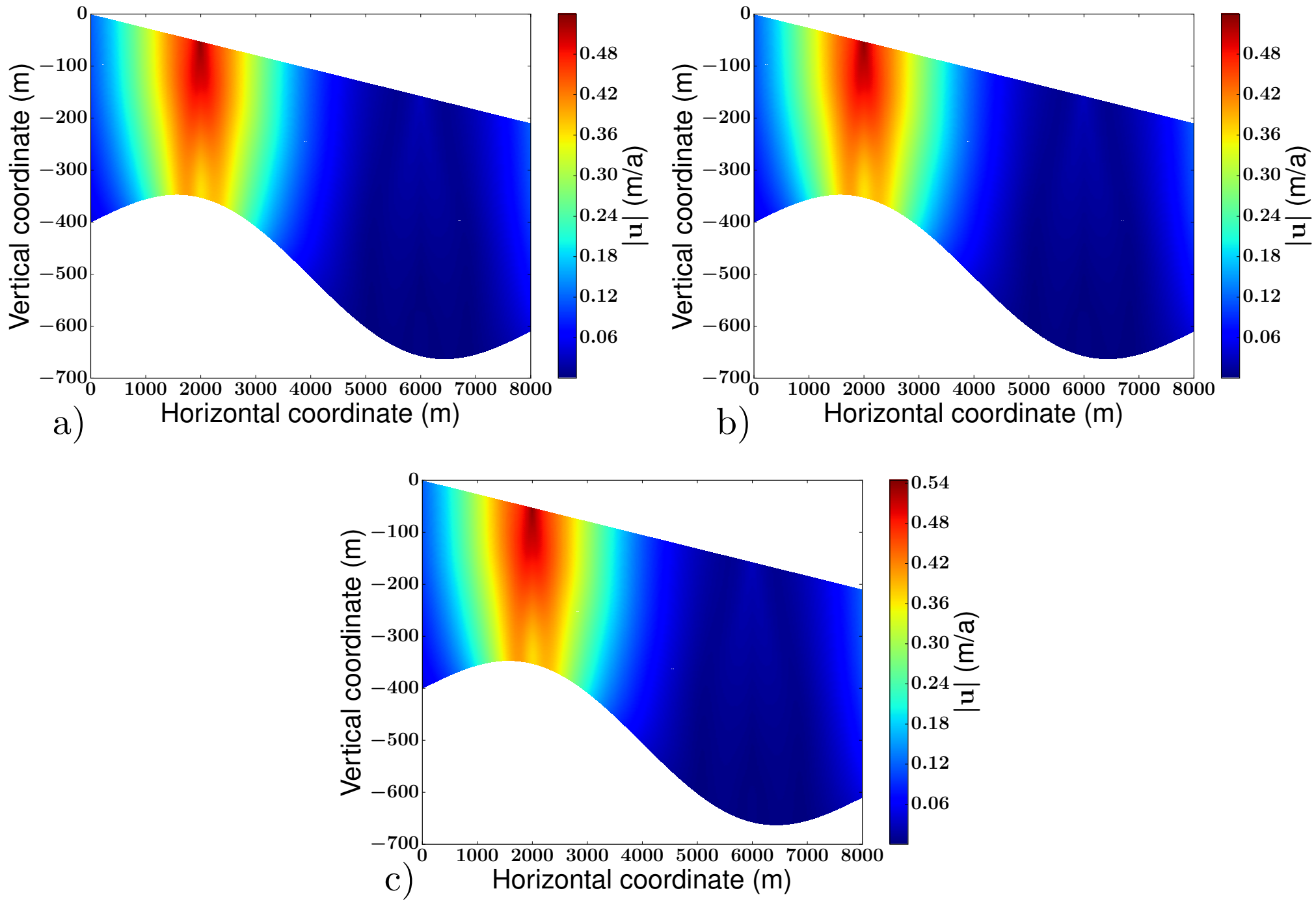

Figure A1. Norm of difference between the manufactured solutions and the (a) strong, (b) weak and (c) approximative methods.

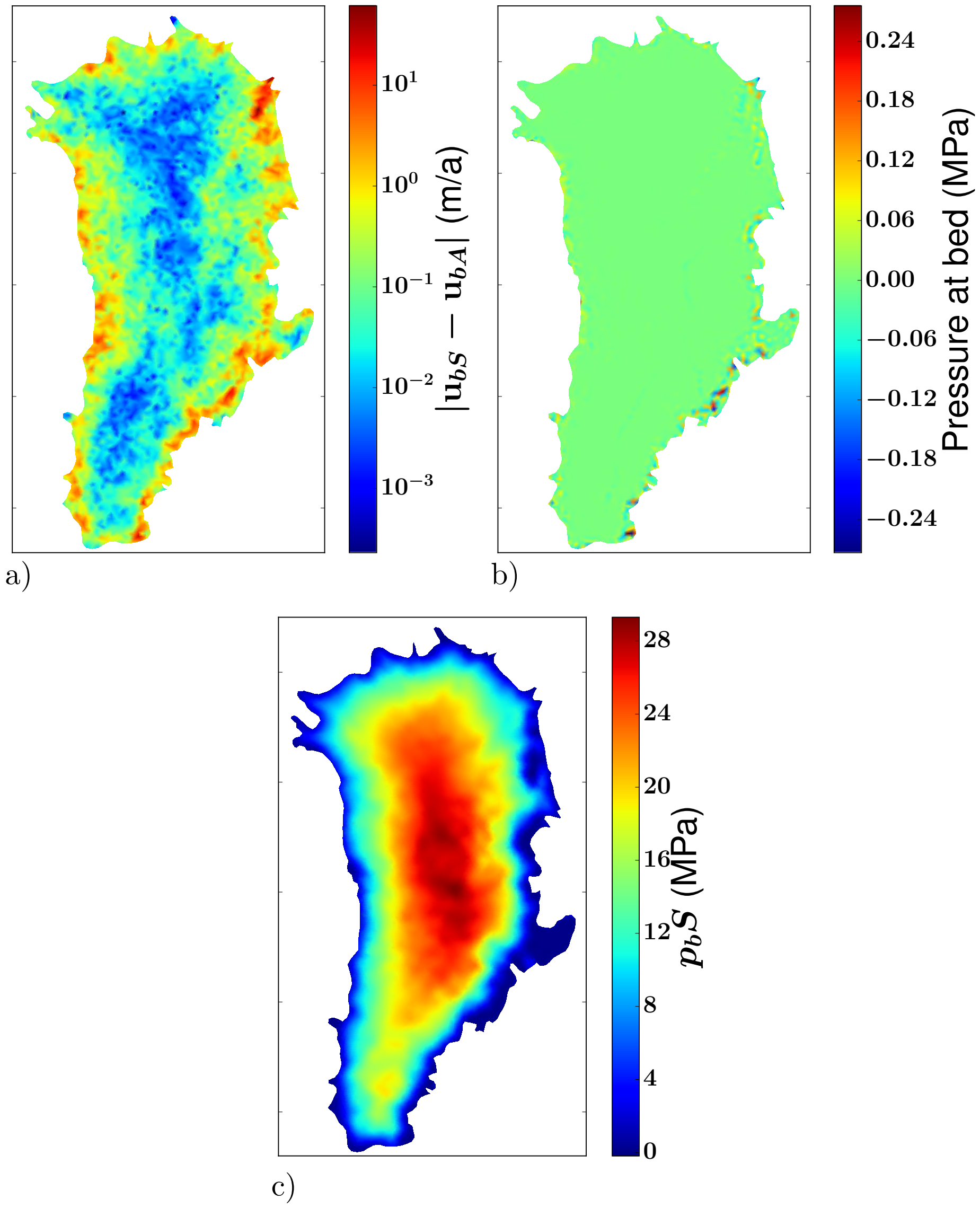

Figure A2. The differences between strong and approximative method shown for (a) basal velocity (norm) and (b) basal pressure (absolute). Panel (c) shows the simulated basal pressure by the strong method.